\RequirePackage{fix-cm}
\documentclass[smallextended]{svjour3}       
\smartqed  

\usepackage{algorithm}
\usepackage{algorithmic}
\usepackage{booktabs} 
\usepackage{upgreek} 
\usepackage{amssymb}
\usepackage{calc}
\usepackage{tikz-cd}
\usepackage{tikz}
\usetikzlibrary{shapes,arrows}
\usepackage{hyperref}
\usepackage{graphicx,verbatimbox}
\usepackage{listings}
\usepackage{wrapfig,lipsum}
\usepackage{subcaption}
\usepackage{mathtools}
\usepackage{color}
\usepackage{tabularx,ragged2e}
\usepackage{multirow}
\usepackage{mathtools}
\usepackage{enumitem}
\usepackage[british]{babel}
\usepackage[numbers]{natbib}
\usepackage{comment}
\usepackage{color}
\usepackage{amsmath,nccmath,tabularx} 
\usepackage[section]{placeins}
\usepackage{natbib}
\let\Oldsection\section
\renewcommand{\section}{\FloatBarrier\Oldsection}

\let\Oldsubsection\subsection
\renewcommand{\subsection}{\FloatBarrier\Oldsubsection}

\let\Oldsubsubsection\subsubsection
\renewcommand{\subsubsection}{\FloatBarrier\Oldsubsubsection}

\newcommand\numberthis{\addtocounter{equation}{1}\tag{\theequation}}

\graphicspath{{images/}}

\begin{document}

\title{Reinforcement Online Learning to Rank with Unbiased Reward Shaping
}


\author{Shengyao Zhuang         \and Zhihao Qiao \and
        Guido Zuccon 
}


\institute{Shengyao Zhuang \at
              The University of Queensland \\
              \email{s.zhuang@uq.edu.au}           
           \and
           Zhihao Qiao \at
           The University of Queensland \\
           \email{zhihao.qiao@uq.edu.au} 
           \and
           Guido Zuccon \at
           The University of Queensland \\
           \email{g.zuccon@uq.edu.au} 
}


\maketitle

\begin{abstract}
Online learning to rank (OLTR) aims to learn a ranker directly from implicit feedback derived from users' interactions, such as clicks. 
Clicks however are a biased signal: specifically, top-ranked documents are likely to attract more clicks than documents down the ranking (position bias). 
In this paper, we propose a novel learning algorithm for OLTR that uses reinforcement learning to optimize rankers: Reinforcement Online Learning to Rank (ROLTR). In ROLTR, the gradients of the ranker are estimated based on the rewards assigned to clicked and unclicked documents. 
In order to de-bias the users’ position bias contained in the reward signals, we introduce unbiased reward shaping functions that exploit inverse propensity scoring for clicked and unclicked documents. The fact that our method can also model unclicked documents provides a further advantage in that less users interactions are required to effectively train a ranker, thus providing gains in efficiency.
Empirical evaluation on standard OLTR datasets shows that ROLTR achieves state-of-the-art performance, and provides significantly better user experience than other OLTR approaches. To facilitate the reproducibility of our experiments, we make all experiment code available at \url{https://github.com/ielab/OLTR}. 
\end{abstract}

\section{Introduction} \label{intro}
Learning to rank (LTR) is a supervised machine learning technique that has been widely used in modern search engines to learn rankers. \textit{Explicit feedback}, consisting of  assessors manually judging the relevance of query-document pairs, is required for LTR~\cite{liu2011learning,li2011learning}. 
This labelled dataset requirement poses obvious limitations: these datasets take substantial effort and cost to compile~\cite{qin2010letor,qin2013introducing,chapelle2011yahoo}, labelling personal documents is unethical and often impossible~\cite{wang2016learning}, static datasets cannot model user intent change over time~\cite{lefortier2014online,zhuang2021how}, and user preferences may not agree with that of annotators~\cite{sanderson2010test}.

In order to overcome these limitations, \textit{implicit feedback}, such as clicks, has been leveraged.
This type of training signal is not affected by the above limitations and has been an attractive alternative to annotated datasets~\cite{joachims2002optimizing}. However, training rankers with implicit feedback has its own drawbacks and challenges. For example, clicks are a weak relevance signal because they often are affected by a number of biases and noise. One of the most prominent bias in web search is the \textit{position bias}, where higher ranked documents have a higher chance to be observed and thus gain more clicks, even if they may be not relevant~\cite{guan2007eye,pan2007google,hofmann2016online,joachims2017unbiased}. Therefore, it is important to consider the influence of such biases.

Two main families of approaches have emerged that attempt to learn effective rankers from users' implicit feedback~\cite{jagerman2019model}: \textit{counterfactual learning to rank} (CLTR) \cite{joachims2017unbiased} and \textit{online learning to rank} (OLTR) \cite{yue2009interactively}. 
In CLTR, given a historical click through log, clicks are treated as pure binary relevance labels and \textit{inverse propensity scoring} (IPS) is used to re-weight clicks in order to discount the effect of biases. Rankers are trained in an offline manner and deployed online after training. This offline batch updates pipeline can avoid the risk of exposing users to low-quality results since it only displays the best search engine results that are possible for a given CLTR algorithm and training data.

On the other hand, OLTR algorithms interactively update rankers after each user interaction has taken place, thus being more responsive to a non-stationary user environment~\cite{zhuang2021how}.  In contrast to CLTR, current OLTR methods do not directly model position or selection bias, and only assume relevant documents are more likely to be clicked than non-relevant documents~\cite{joachims2002optimizing}. The biases and noise of users clicks are handled by \textit{online interventions}~\cite{jagerman2019model}, i.e., slightly perturbated result lists are displayed and preferences towards rankers are informed by users clicks. This is one of the key aspects of OLTR, but also one of its biggest disadvantages that limits OLTR's uptake in practice: such online interventions carry the risk of displaying a ``sub-optimal'' ranking list directly to the user, thus hurting user experience (as measured by online evaluation metrics). Hence, it is a requirement for OLTR methods to efficiently leverage the click feedback so that a good ranker is learnt as fast as possible to avoid displaying low quality search results to a large number of users.

To move beyond the limitations of existing OLTR methods, in this article, we propose a novel OLTR algorithm called \textit{Reinforcement Online Learning to Rank} (ROLTR), which exploits the reinforcement learning (RL) approach adapting it to OLTR. Our motivation for using RL in OLTR is based on the following observations: 
\begin{enumerate}
	\item RL suits OLTR setting very well: RL is powerful for modeling interactive environments and maximizing the long-term rewards yield from the environment. In the OLTR setting, the interactive environment is composed by the users and the search system, and the rewards are the users’ satisfaction.
	\item RL with carefully designed reward functions allows OLTR algorithms to directly remove the biases present in the user clicks: this is currently hard for other OLTR algorithms to achieve. 
\end{enumerate}

As we show in this article, our proposed ROLTR can directly remove position bias and thus effectively and, importantly, efficiently (i.e. within less impressions) update the ranker. To achieve this, we formalize OLTR as a \textit{Markov Decision Process} (MDP) problem and use \textit{policy gradient} with rewards assigned on clicked and unclicked documents to estimate the update gradients. In order to de-bias users' clicks, we further introduce unbiased reward shaping functions that re-weight the rewards for both clicked and unclicked documents. We mathematically prove that the gradient estimation of ROLTR is unbiased with respect to position bias, and it can directly optimize IR metrics such as discounted cumulative gain (DCG). The idea of leveraging unclicked data has been recently explored in offline counterfactual learning studies~\cite{hu2019unbiased,wang2021non}, however, our method is currently the only OLTR approach that can gain unbiased learning signals from unclicked documents, thus speeding up convergence. Empirical results further show that ROLTR significantly outperforms traditional OLTR methods and is at par with current state-of-the-art methods (offline performance), although requiring fewer user interactions. As a result, our method delivers considerably better user experience (online performance).
\section{Related work}

\subsection{Counterfactual Learning to Rank}
Unlike traditional LTR where rankers are learned from explicitly labelled datasets~\cite{liu2011learning}, counterfactual LTR~\cite{agarwal2018counterfactual,ai2018unbiased,joachims2017unbiased} uses historical interaction data, typically click logs, to learn a ranker. However, clicks are a biased signal. The most prominent bias in the click signal is the \textit{position bias}: assuming that users examine search engine result pages (SERPs) from top to bottom, then the results that are ranked higher are more likely to be observed by the users~\cite{joachims2017unbiased}. Joachims et al. refer to the probability of a search result at a rank $i$ to be observed as its \textit{propensity}~\cite{joachims2017unbiased}. They then define the \textit{inverse propensity scoring} (IPS) method to re-weight user clicks: when IPS is used, the estimated ranking score is unbiased with respect to position bias. One crucial requirement of IPS-based CLTR methods is the prior knowledge of user propensity. This is usually estimated by conducting online result randomization which can negatively affect the user experience. To address this issue,  Ai et al.~\cite{ai2018unbiased} proposed the Dual
Learning Algorithm (DLA) to jointly learn an unbiased ranker
and an unbiased propensity model, thus avoiding the preprocessing of propensity estimations. Unlike these CLTR works, our method also uses IPS but in the context of reinforcement learning for OLTR and we further propose a new IPS method for unclicked documents so as to gain an unbiased training signal from unclicked documents. This accelerates the learning process because both click and unclick information is used for training, thus increasing the number of implicit signals obtained from each search and used for training.


In addition to position bias, recent work in counterfactual LTR has considered correcting the \textit{selection bias}~\cite{ovaisi2020correcting,oosterhuis2020policy}, in which some documents have zero probability of being observed by the users (thus never having a chance to be clicked and contribute a training signal). We note that the selection bias is fundamentally different from the position bias in terms of that this bias is introduced by the system itself: some documents will never be included by the system in the top SERPs. In contrast, the position bias instead comes from the users. This often occurs in web search where SERPs only show a small subset of documents in the first page (e.g., 10) and users do not proceed beyond the first SERP. In this circumstance, documents ranked beyond the first SERP have no chance of being observed, hence never get identified as positive training examples. To debias the selection bias in the click signals, Oosterhuis and Rijke~\cite{oosterhuis2020policy} proposed a policy-aware counterfactual estimator for CLTR to directly account for the selection bias introduced by a stochastic logging policy. On the other hand,  Ovaisi et al.~\cite{ovaisi2020correcting} adapted Heckman’s two-stage method
to account for selection and position bias in LTR systems. In our work, we follow the standard OLTR and CLTR experimental setup and simulate typical, real-world circumstances; we also include selection bias in our experiments by only placing 10 documents in the SERPs. Although in this paper we directly focus on correcting position bias, selection bias is also partially corrected by the proposed method via online intervention. Nevertheless, recently proposed methods~\cite{ovaisi2020correcting,oosterhuis2020policy} could be used to further modify our reward shaping functions: this would have the effect of further reducing the selection bias of the gradient estimations.

\subsection{Online Learning to Rank}
\begin{figure}[t]
	\includegraphics[width=\linewidth]{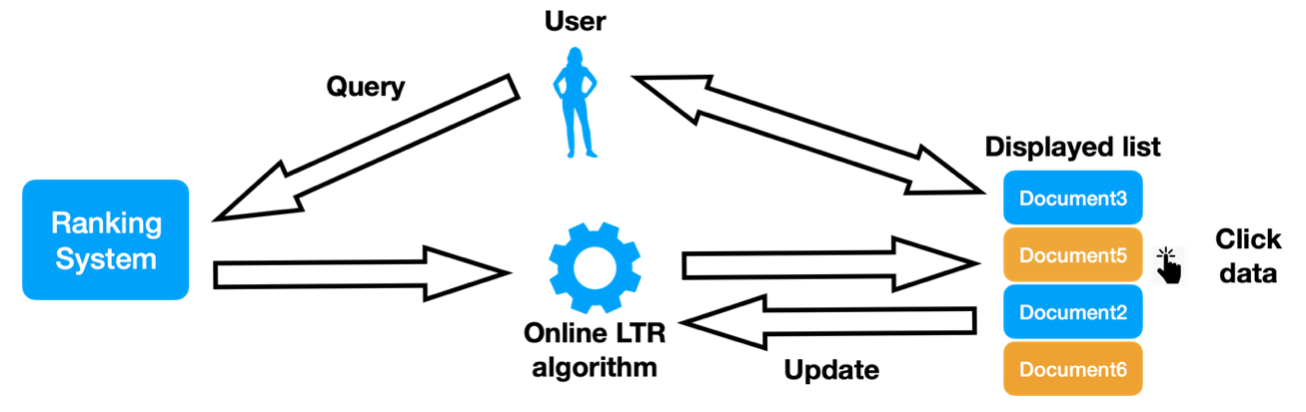}
	\caption{A schematic view of the online learning to rank process. Users pose queries to the ranking system. This uses a online LTR algorithm to identify the candidate documents to display to the user. Depending on the OLTR algorithm used, candidate documents will be used to form a displayed list. The user examines the displayed search results (SERP) and clicks on items of interest. The click feedback is used by the OLTR algorithm to perform ranker updates.}
	\label{OLTR-process}
\end{figure}

Similarly to CLTR, online LTR (OLTR) also considers implicit user feedback to learn a ranker. Unlike CLTR, however, this is done online, by directly interacting with users. This online training process allows to control data acquisition and handle biases and noise through online interventions with regards to which documents to display. Figure~\ref{OLTR-process} provides a schematic representation of the OLTR process.

The \textit{Dueling Bandit Gradient Descent} (DBGD)~\cite{yue2009interactively} models OLTR as a dueling bandit problem and uses \textit{online evaluation} with users' clicks on interleaved or multileaved SERPs to indicate user preference among a pool of candidate rankers~\cite{schuth2015probabilistic,oosterhuis2016probabilistic,hofmann2011balancing,hofmann2013reusing,schuth2016multileave,hofmann2011probabilistic}. More recent work has studied methods to reduce the variance of the gradients estimated by online evaluation~\cite{wang2019variance,wang2018efficient}.
 However, because gradients are updated towards the winning rankers in the candidate pool, the quality of the gradient estimation is influenced by the number of candidate rankers. When the number of candidate rankers is large, the performance of DBGD is limited in terms of both effectiveness and efficiency~\cite{schuth2015probabilistic,li2020mergedts}. The recently proposed \textit{Counterfactual Online Learning to Rank} (COLTR)~\cite{zhuang2020counterfactual} attempts to overcome the issues associated with online evaluation. This is done by replacing online evaluation with counterfactual evaluation for DBGD. Instead of interleaving, COLTR uses clicks collected by the current ranker to evaluate candidate rankers thus providing high efficiency. However, empirical results have shown COLTR requires more exploration, thus possibly hurting user experience (online evaluation metrics).

Unlike DBGD and COLTR, the \textit{Pairwise Differentiable Gradient Descent} (PDGD) \cite{oosterhuis2018differentiable} does not require to sample candidate rankers for online evaluation. Instead, PDGD directly estimates gradients based on pairwise preferences between documents in the SERP, inferred by users' clicks; then, stochastic gradient descent is used to update the ranker. The gradient estimation of PDGD is unbiased with respect to user document pair preferences~\cite{oosterhuis2018differentiable}. PDGD is empirically found to be significantly better than DBGD in terms of final convergence, learning speed and user experience during optimization, making PDGD the current state-of-the-art method for OLTR~\cite{jagerman2019model,zhuang2020counterfactual,oosterhuis2019optimizing,wang2021federated}. PDGD has also been adapted to the federated OLTR context~\cite{wang2021effective}, exhibiting again state-of-the-art performance.  However, both PDGD and DBGD based methods do not directly use IPS to handle position bias, which has been proven to be very important in previous CLTR results~\cite{joachims2017unbiased}.

As mentioned above, CLTR and OLTR differ from the setting considered for learning the ranker: use an historical log of implicit interactions for counterfactual, versus use direct observation and interaction with rankings in the case of online LTR. We note that Jagerman et al.~\cite{jagerman2019model} have specifically studied the similarity and differences between counterfactual and online LTR. They suggested that, from a theoretical standpoint, counterfactual LTR exploits position bias better, however they also indicated that empirical results have shown that OLTR (and in particular PDGD) is more reliable.  On the other hand, recent works have been focusing on adapting the offline counterfactual learning to the online setting~\cite{ai2021unbiased,zhuang2020counterfactual,oosterhuis2020taking,oosterhuis2021unifying}. These works have suggested that OLTR algorithms can benefit from the counterfactual learning framework. Based on this context, our proposed method can also be thought of as embracing this direction.

\subsection{Reinforcement Learning to Rank}
Reinforcement learning (RL) has previously been applied to offline LTR problems, but not to online LTR. Wei et al.~\cite{wei2017reinforcement} have formalized ranking as a Markov Decision Process (MDP) problem and introduced the MDPRank algorithm to optimize a linear ranker. Specifically, the document ranking is modelled as a sequential decision making process where each time step corresponds to a position in the ranking and the action taken at each time step corresponds to the selection of a document for the corresponding position. The rewards given by the environment are generated according to the relevance label of the documents, and the classic policy gradient algorithm of REINFORCE~\cite{sutton2000policy} is used to maximize the expectation of cumulative rewards received by the ranker. An attractive property of MDPRank is that it can use gradient descent to directly optimize non-differentiable ranking metrics such as DCG. However, Xu et al.~\cite{jun2020ppg} have recently shown that the gradient estimation of the original MDPRank exhibits high variance, and thus this method requires more training episodes to learn an effective ranker. A similar reinforcement learning framework has also been used for search result diversification~\cite{feng2018greedy,xia2017adapting}, multi-page search~\cite{zeng2018multi} and recommendation system~\cite{zhao2018deep,zhao2018recommendations,zhao2021dear}.

In this paper, we also apply a MDPRank-like algorithm, although in the context of OLTR and with two important adaptions. First, the original MDPRank is designed for traditional offline LTR, i.e., the relevance labels of query-document pairs are provided. In the OLTR setting, however, only biased user interaction data such as clicks is available: thus the rewards given by the online environment are biased. In order to obtain unbiased rewards, we introduce unbiased reward shaping functions for MDPRank to discount position bias. Secondly, the high variance of the gradient estimation of MDPRank makes it converge slowly. However, for OLTR to be viable, we need the algorithm to converge fast so that it does not hurt the user experience too much. To reduce variance, we simplify the objective function of the policy gradient used in the original MDPRank to achieve an objective function with much lower variance of gradient estimation, without changing its optimization target.

\section{Method}

\subsection{Ranking as a Markov Decision Process for Online Learning to Rank}
Ranking can be formalized as an MDP problem, where the search engine (agent) has to decide which document to place (action) at rank $i$, given the current candidate document set (state)~\cite{wei2017reinforcement}. In the offline LTR setting, i.e., when relevance labels are provided, any IR evaluation measure can be used as a reward function for the ranking created by the search engine. In the online LTR setting, however, rewards are typically inferred via users‘ implicit feedback, such as clicks. Following Wei et al.~\cite{wei2017reinforcement}, who have limited their attention to offline LTR only, we define each component of the MDP ranking for the OLTR as follows: 

\textbf{States} $\boldsymbol {S}$ indicates the set of states that the agent (search engine) will observe from the environment. In the MDP ranking, at a time step $t$, a state $s_t\in{S}$ is the candidate document set $D_t$ which contains the documents that need to be ranked.

\textbf{Actions} $\boldsymbol {A}$ is the set of possible actions that the agent can take when at a state $s_t$. An action at a time step $t$, denoted as $a_t\in{A(s_t)}$, consists of selecting a document $d_{m(a_t)}\in{D_t}$ to place at rank $i$, where rank $k = t+1$\footnote{Note that the time step $t$ starts from 0 and the rank position $k$ starts from 1.}, $m(a_t)$ is the function that maps the action to the document index, and $d_{m(a_t)}$ is represented by a feature vector. We note the feature elements in $d_{m(a_t)}$ are not only obtained based on the document but also on the query.

\textbf{Transition} $\boldsymbol {T(s,a)}$ is the function $T(s_t,a_t) = s_{t+1}$, which maps a state $s_t$ to its next state $s_{t+1}$ based on the action $a_t$, where $s_{t+1}$ is the candidate document set $D_t$ without the selected document $d_{m(a_t)}$:
\begin{equation} \label{transaction}
	s_{t+1} = s_t \setminus d_{m(a_t)} =D_t \setminus d_{m(a_t)} = D_{t+1}
\end{equation}

 \textbf{Policy} $\boldsymbol {\pi(a|s, \theta)}$ is a probability distribution over all possible actions that the agent can take when in the state $s$, given the current model parameters $\theta$ (where $\theta$ is a vector containing the model's parameters). We compute this distribution using the softmax function over document scores:
\begin{equation}\label{policy}
	\pi(a_t|s_t, \theta) = 
	\frac{exp \left( f_\theta(d_{m(a_t)} \right) }
	{\sum_{a\in A(s_t)}exp \left( f_\theta(d_{m(a_t)} \right) }
\end{equation}

$f_\theta(d_{m(a_t)})$ is the document relevance score estimated by the ranking model. In our experiments, we use a linear ranking model, this means the dimensionality of $\theta$, i.e., the number of parameters in the ranking model, is equal to the dimensionality of the feature vector. However, we note that the ranking model can be extended to any dimensionality, e.g., a neural ranking model. For simplicity, we use $a_t \sim \pi(s_t)$ to denote the action $a_t$ selected according to the policy $\pi$ at state $s_t$.
 
 \textbf{Reward} $\boldsymbol {R(s,a)}$ is the reward function that models the immediate reward given by the environment and its value represents the reward assigned to the action $a$ that has been selected by the agent in state $s$. 
 When full information is provided (i.e., the true relevance labels are given), it is straightforward to define the reward function on the basis of IR evaluation measures such as DCG. However, in the OLTR setting, relevance labels are not provided -- in place of these, the ranker observes implicit, noisy and biased user feedback such as clicks. Thus, in such a partial information setting, the reward function is defined based on the implicit feedback, e.g., on click labels. We define the value of the reward function next.


\subsection{Unbiased reward shaping}\label{unbiasedness}
In RL, reward shaping is used to reshape the original reward function to better guide the direction of the gradient update~\cite{ng1999policy}. Prior knowledge about the environment is needed to formalize a reliable reward shaping function to avoid otherwise to bias learning~\cite{popov2017data}. 

\subsubsection{Na\"ive reward function}

In the (offline) LTR approach by Wei et al.~\cite{wei2017reinforcement}, DCG scores have been used as the original reward function to instruct the search engine to learn a ranker, when true relevance labels are supplied. This reward function  $R_{DCG}(s_t,a_t)$ is defined as:

  \begin{equation} \label{eq:3}
R_{DCG}(s_t,a_t) = \lambda(t)\cdot y_{m(a_t)}
\end{equation}

\noindent where $\lambda(t)=\frac{1}{log_2(t+2)}$ is the DCG weight term, and $y_{m(a_t)}\in\{0,1\}$ is the binary relevance label of the document at rank $t+1$. This is a reasonable reward function, provided that relevance labels are known. This reward function directly corresponds to the DCG evaluation metric (because of the DCG weight term), and thus the agent will attempt to learn a policy $\pi$ that directly maximizes the cumulative rewards $\Delta$ (DCG scores) of the ranking episode, for any given initial state $s_0$:

  \begin{equation} \label{eq:5}
\Delta_{R_{DCG}}( s_0, \pi, y) = \sum_{t=0}R_{DCG}(s_t,\pi(s_t)) 
= \sum_{y_{m(a_t)}=1}\lambda(t)
\end{equation}

However, in the \textit{online} LTR setting, relevance labels are unknown (i.e., this is a partial information setting); instead, users' clicks are used as learning signal. A na\"ive reward function in this context is then to treat clicks as binary relevance labels and thus use the DCG reward function of equation~\ref{eq:3}: 

\begin{equation} \label{eq:6}
R_{NAIVE^+} (s_t,a_t) = \lambda(t)\cdot c_{m(a_t)}
\end{equation}

\noindent where $c_{m(a_t)}=1$ represents that a user has clicked on the document at rank $t+1$, while $c_{m(a_t)}=0$ represents no click on the document. If we assume that the probability of a document $d_{m(a_t)}$ to be observed by a user (known as propensity) only depends on the rank position $t+1$ (note that $t$ starts from 0) and no click noise is present, i.e., $P(o_{t+1}=1 | t+1)$, then the expectation of the final ranking rewards is biased to the users' propensity~\cite{joachims2017unbiased,jagerman2019model}:
  
  \begin{equation} \label{eq:7}
   \begin{aligned}
\mathbb{E}_o[\Delta_{R_{NAIVE^+}}(s_0, \pi, c) ] &=\sum_{t=0}R_{NAIVE^+}(s_t,\pi(s_t))  \\
&= \sum_{y_{m(a_t)}=1}P(o_{t+1}=1 | t+1) \cdot \lambda(t) 
\end{aligned}
\end{equation}

This is the so-called position bias effect: top-ranked documents usually have a larger chance of being observed, and thus the rewards assigned to them are biasedly higher.

\subsubsection{IPS reshaping}
Recent work in CLTR has attempted to account for this position bias effect~\cite{ai2018unbiased,joachims2017unbiased}. One of the most commonly used approaches to mitigate position bias is \textit{Inverse Propensity Scoring} (IPS). We follow this direction to unbias the click signal in our RL framework, and define the IPS reward shaping function as $F_{IPS^+}$ and the reshaped reward function as $R_{IPS^+}$:

\begin{equation} 
  \begin{aligned}
R_{IPS^+} (s_t,a_t)& = F_{IPS} (R_{NAIVE^+} (s_t,a_t))\\
&= \frac{R_{NAIVE^+}(s_t,a_t)}{P(o_{t+1}=1 | t+1)}= \frac{\lambda(t)}{P(o_{t+1}=1 | t+1)} \cdot c_{m(a_t)}
\end{aligned}
\end{equation}

It can be proven that the cumulative na\"ive rewards reshaped by $F_{IPS^+}$ are an unbiased estimate of the ranking rewards with true relevance labels~\cite{joachims2017unbiased}: $\mathbb{E}_o[\Delta_{R_{IPS^+}}(s_0, \pi, c) ] =\Delta_{R_{DCG}}(s_0, \pi, y)$. 
%
%
Thus, $F_{IPS}$ can be used to obtain unbiased cumulative rewards with respect to position bias, likely providing more reliable gradient estimations than the na\"ive reward function.

\subsubsection{Negative rewards}
The reward functions above only provide positive rewards to the clicked/relevant documents, while they assign a zero reward to the unclicked/irrelevant documents. However, in other RL problems, it is often found that negative rewards help the agent to avoid selecting poor actions~\cite{sutton2018reinforcement}: for a ranker, negative rewards can help avoid selecting irrelevant documents from the document set, thus leading to better user experience. Based on this, we introduce  a negative DCG reward function for online LTR (a similar one could be defined for offline LTR): 

\begin{equation}
R_{NAIVE^-} (s_t,a_t) = \lambda(t)\cdot (c_{m(a_t)}-1)
\end{equation}

Hence, the negative cumulative ranking score is calculated by:

\begin{equation}
\Delta_{R_{NAIVE^-}} (s_0, \pi, c) = \sum_{t=0}R_{NAIVE^-}(s_t,\pi(s_t)) = \sum_{c_{m(a_t)}=0}-\lambda(t)
\end{equation}

This means highly ranked unclicked documents will be penalised more and thus drag down the final ranking reward. It can be proven that this na\"ive negative reward function is a biased estimate of the negative DCG score of the ranking ($\Delta_{R_{DCG^-}} ( s_0, \pi, y)$):

\begin{align*}
&\mathbb{E}_o[\Delta_{R_{NAIVE^-}} (s_0, \pi, c)] \\
&= \mathbb{E}_o\left[\sum_{t=0}R_{NAIVE^-} (s_t,\pi(s_t))\right]
=\mathbb{E}_o\left[\sum_{c_{m(a_t)}=0}-\lambda(t)\right] \\ 
&=\mathbb{E}_o\left[\sum_{ y_{m(a_t)}=0}-\lambda(t)\right]  + \mathbb{E}_o\left[\sum_{o_{t+1}=0\wedge y_{m(a_t)}=1}-\lambda(t)\right] \\ 
&= \sum_{y_{m(a_t)}=0}-\lambda(t) - \sum_{y_{m(a_t)}=1}(1-P(o_{t+1}=1|t+1))\cdot\lambda(t)\\
&=\Delta_{R_{DCG^-}} ( s_0, \pi, y) - \sum_{y_{m(a_t)}=1}(1-P(o_{t+1}=1|t+1))\cdot\lambda(t) \numberthis \label{eqn}
\end{align*}

That is, the na\"ive negative reward overestimates the true negative DCG  reward by counting the relevant but not observed documents as irrelevant. 

\subsubsection{Negative IPS reward shaping function}
In order to get unbiased negative DCG rewards, we propose the negative IPS reward shaping function $F_{IPS^-}$ and denote the reshaped negative  DCG reward as $R_{IPS^-}$:

\begin{align*}
&R_{IPS^-} (s_t,a_t)=F_{IPS^-}(R_{NAIVE^-} (s_t,a_t)) \numberthis \label{eqn}\\
& = \lambda(t)\cdot (c_{m(a_t)} - 1) + \left (\frac{1-P(o_{t+1}=1|t+1)}{P(o_{t+1}=1|t+1)} \right )\lambda(t)\cdot c_{m(a_t)}
\end{align*}

We mathematically prove that $\mathbb{E}_o[\Delta_{R_{IPS^-}} (s_0, \pi, c)]$ is an unbiased estimate of $\Delta_{R_{DCG^-}} (s_0, \pi, y)$:
	\begin{align*}
	&\mathbb{E}_o[\Delta_{R_{IPS^-}} (s_0, \pi, c)] = \mathbb{E}_o\left[\sum_{t=0}F_{IPS^-}(R_{NAIVE^-}(s_t,\pi(s_t)))\right] \\ 
	&=\mathbb{E}_o\left[\sum_{t=0}\lambda(t)\cdot (c_{m(a_t)} - 1) + \left (\frac{1-P(o_{t+1}=1|t+1)}{P(o_{t+1}=1|t+1)} \right )\lambda(t)\cdot c_{m(a_t)}\right] \\
	&=\mathbb{E}_o\left[ \sum_{t=0}\lambda(t)\cdot (c_{m(a_t)} - 1)\right] +\\
	&+\mathbb{E}_o\left[ \sum_{t=0}\left(\frac{1-P(o_{t+1}=1|t+1)}{P(o_{t+1}=1|t+1)} \right )\lambda(t)\cdot c_{m(a_t)}\right] \\
	&=\mathbb{E}_o[\Delta_{R_{NAIVE^-}} (s_0, \pi, c)]  + \\
	&+\mathbb{E}_o\left[\sum_{o_{t+1}=1\wedge y_{m(a_t)}=1}\left(\frac{1-P(o_{t+1}=1|t+1)}{P(o_{t+1}=1|t+1)} \right )\lambda(t)\right] \\
	&=\Delta_{R_{DCG^-}} (s_0, \pi, y) - \sum_{y_{m(a_t)}=1}(1-P(o_{t+1}=1|t+1))\cdot\lambda(t) \\
	& +\sum_{y_{m(a_t)}=1}(1-P(o_{t+1}=1 | t+1)) \cdot \lambda(t)\\ 
	&=\Delta_{R_{DCG^-}} (s_0, \pi, y)
	\numberthis \label{eqn}\\
	\end{align*}
This reward shaping function allows to assign unbiased negative rewards to unclicked documents so as to obtain an unbiased cumulative negative DCG ranking score. 

\subsubsection{Prior knowledge of propensity}
All the unbiased reward shaping functions above require to know a priori the users' propensities $P(o_{t+1}=1|t+1)$. Many recent works have considered  estimating such propensities from historical click-logs~\cite{agarwal2017effective,fang2019intervention,agarwal2019estimating} and during training~\cite{ai2018unbiased,wang2018position}. We regard propensity estimation as being beyond the scope of this article and in our experiments we assume the propensities to be known. Nevertheless, for completeness, we also test how sensitive our method is to propensity mismatch (Section~\ref{sec_propensity_missmatch}).

\subsection{Learning with policy gradient} \label{policy_gradient}
Following previous work~\cite{wei2017reinforcement,yao2020rlper}, we learn the policy model parameters $\theta$ with REINFORCE~\cite{williams1992simple,sutton2018reinforcement,sutton2000policy}, a widely used policy gradient algorithm. In REINFORCE, the objective is to find an optimal policy that can maximize the expectation of cumulative reward from the beginning of each episode, $J(\theta) = \mathbb{E}_\pi\left[ G_t\right]$, where $t=0$.
Here $G_t$ is the discounted future cumulative reward starting from time step $t$, $G_t = \sum_{m=t}^{M}\gamma^{m-t}\cdot R(s_{m},a_{m})$, and 
$M$ is the maximum depth of the ranking episode, and $\gamma \in [0, 1]$ is the discount factor. In previous offline LTR work~\cite{wei2017reinforcement,jun2020ppg}, $\gamma$ has been set to $1$ (maximum value), which results in $G_0 = \Delta_{R_{DCG}}$. This value results in the learning algorithm directly maximizing DCG. However, it is well known in RL that larger $\gamma$ values will lead the agent to care more about future rewards but at the same time to produce gradient estimations with significantly high variance, thus slowing down the learning speed~\cite{jun2020ppg,sutton2018reinforcement}. In OLTR settings, however, learning speed is very important as it is highly entangled with user experience, as measured by the online performance of the ranker. 

Hence, in order to reduce variance and speed up learning, we simplify MDP to Contextual Bandits~\cite{jagerman2020safe,adomavicius2005incorporating,hofmann2011contextual} by setting $\gamma=0$. This setting makes REINFORCE to choose $a_t$ so as to maximize only the expectation of immediate reward $R(s_t, a_t)$:
 
\begin{equation}
	J(\theta) = \mathbb{E}_\pi\left[ R(s_t, a_t)\right]
\end{equation} 

This setting actually does not change the fact that the objective is to optimize DCG. This is because maximizing the expected immediate reward at a time step $t$ is equivalent to selecting the most likely relevant document from the candidate set $D_t$ for the state $s_t$: this is guaranteed to have maximum expected cumulative reward for that ranking episode. Thus, it is safe to ignore future rewards in this case. In Section~\ref{sec_discount}, we empirically show that $\gamma=0$ enjoys a much faster learning speed than other settings, without any loss in final convergence. It is important to note that the above is not true for tasks such as search result diversification, since greedily choosing the most relevant document at each rank position may cause lower final ranking scores~\cite{feng2018greedy}. Thus, we use the full MDP algorithm for this special situation.

 Following the standard policy gradient practice~\cite{wei2017reinforcement,sutton2018reinforcement}, we estimate gradients with \textit{Monte Carlo sampling} and the gradient $\nabla_\theta J(\theta)$ can be calculated as:
\begin{equation}\label{PG}
\nabla_\theta J(\theta) = R(s_t, a_t)\nabla_\theta log(\pi(a_t|s_t, \theta) )
\end{equation}
Intuitively, the gradients will update the ranker parameters towards the actions that yield the highest immediate reward $R(s_t, a_t)$.


The complete procedure of ROLTR is described in Algorithm~\ref{algo:ReOLT}. At each ranking episode $i$, i.e., at each round of user interaction (line \ref{lst:line:2} for-loop), the search engine receives a query $q_i$ and the initial candidate document set $D_0$ is generated (line \ref{lst:line:3}). Then the algorithm first draws an action (document) from the distribution created by the current policy (line \ref{lst:line:6}). Then, the selected action, the state information, and the corresponding document are recorded (line \ref{lst:line:7}) and the environment moves to the next state (line \ref{lst:line:8}). The same procedure is repeated  for the next rank positions, until the algorithm reaches the maximum depth for the ranking (line \ref{lst:line:5} for-loop). After finishing ranking, the final result list is shown to the user, who provides feedback to the search engine in the form of click labels (line \ref{lst:line:10}). Next, for each recorded state-action pair (line \ref{lst:line:11} for-loop), the gradient is calculated (line \ref{lst:line:14}) and the current policy is updated at the end of each ranking episode (line \ref{lst:line:16}).

	\begin{algorithm}[t]
	\caption{ Reinforcement Online Learning to Rank (ROLTR).}
	\begin{algorithmic}[1]
		\STATE {\bf Input}: Initial weights $\theta$, learning rate $\alpha$, reshaped reward function $R$, transition function $T$, number of docs on SERPs $M$.
		\FOR{$i$ $\leftarrow$ 0....$\infty$} \label{lst:line:2}
		\STATE $q_{i} \leftarrow receive\_query(i)$, $D_{0}\leftarrow canditate\_set(q_i)$ \label{lst:line:3}
		\STATE $s_0 \leftarrow D_0$, $L_{i}\leftarrow [$ \space $]$, $\Delta \theta \leftarrow 0$ 
		\FOR{$t$ $\leftarrow$ 0....$M$} \label{lst:line:5}
		\STATE $a_t \leftarrow \pi(s_t, \theta)$\hfill// Eq.\ref{policy}  \label{lst:line:6}
		\STATE $L_i.append(s_t, a_t, d_{m(a_t)})$ \label{lst:line:7}
		\STATE $s_{t+1} \leftarrow T(s_t,a_t)$\hfill// Eq.\ref{transaction} \label{lst:line:8}
		\ENDFOR
		\STATE $C_i\leftarrow receive\_clicks(L_{i})$ \label{lst:line:10}
		\FOR{$t$ $\leftarrow$ 0....$M$} \label{lst:line:11}
		\STATE $s_t, a_t \leftarrow L_i.get(t)$, $c_{m(a_t)}\leftarrow C_i.get(t)$ 
		\STATE $\Delta \theta \leftarrow \Delta \theta+ R(s_t, a_t)\nabla_\theta log(\pi(a_t|s_t, \theta) )$ \hfill// Eq.\ref{PG} \label{lst:line:14}
		\ENDFOR
		\STATE $\theta \leftarrow \theta + \alpha\Delta \theta$ \label{lst:line:16}
		\ENDFOR
	\end{algorithmic}
	\label{algo:ReOLT}
\end{algorithm}

\subsection{Guarantee of Unbiasedness for ROLTR}
ROLTR has two main parts: the learning part (Section~\ref{policy_gradient}), which uses the REINFORCE algorithm, and the unbiased reward shaping part (Section~\ref{unbiasedness}).  The theoretical guarantees of REINFORCE are well studied in  reinforcement learning, including the guarantee that the gradient estimate is unbiased with respect to maximizing the expected rewards~\cite{sutton2018reinforcement}. However, if the rewards are biased (which are when the naive reward function is used), then the learning algorithm is biased. This is where the second part of ROLTR comes into play: the unbiased rewards shaping functions guarantee the reward signals assigned to REINFORCE are unbiased with respect to position bias (mathematical proof in Eq.~\ref{eqn}). Thus, the update gradients estimated by ROLTR are guaranteed to maximize unbiased expected rewards, and thus the gradient estimation is unbiased with respect to position bias.

\section{Empirical Evaluation}
To study the effectiveness of ROLTR, we designed a number of empirical experiments aimed to answer the following research questions:

\begin{itemize}[leftmargin=40pt]
	\item[\bf RQ1:] How does the reward discount factor in ROLTR affect gradient variance and final convergence (i.e., the offline nDCG score on the test dataset)?
	\item[\bf RQ2:] How do the unbiased reward shaping functions of ROLTR impact performance?
	\item[\bf RQ3:] How does ROLTR compare in terms of convergence and learning speed against current representative OLTR methods?
	\item[\bf RQ4:] Does ROLTR deliver better user experience than current OLTR methods, i.e., higher online nDCG?	
	\item[\bf RQ5:] How sensitive is ROLTR to  propensity mismatch?
\end{itemize}



\subsection{Datasets and synthetic data generation}
We consider three benchmark datasets that are commonly used to evaluate OLTR~\cite{zhuang2020counterfactual, schuth2015probabilistic,oosterhuis2016probabilistic,oosterhuis2018differentiable,wang2019variance} and CLTR~\cite{jagerman2019model,ai2018unbiased,joachims2017unbiased}: MSLR-WEB10K~\cite{qin2013introducing}, Yahoo! Webscope~\cite{chapelle2011yahoo}, and Istella~\cite{dato2016fast}. MSLR-WEB10K contains 10,000 queries and 125 retrieved documents on average; documents are represented by 136 features. Yahoo! is a bigger dataset, with 29,921 queries and an average of 23.7 documents per query, represented using 700 features. Istella is the largest dataset we consider, with 33,118 queries and an average of 315 documents per query, represented by 220 features. Query-document pair relevance labels for all datasets are recorded on a five-point scale from not relevant (0) to perfectly relevant (4) and have been split into training, validation and test sets (according to the standard splits in the datasets). Queries in the three sets are disjoint. We use the training set to train the rankers, the validation set to tune the hyper-parameters, and the test set to evaluate the rankers' performance.

To avoid hurting user experience, it is common for research in OLTR and CLTR to simulate users' clicks by relying on the relevance labels recorded in the datasets~\cite{zhuang2020counterfactual, schuth2015probabilistic,oosterhuis2016probabilistic,oosterhuis2018differentiable,jagerman2019model,joachims2017unbiased,vardasbi2020cascade}. This also allows to fully control users' biases and noise so that algorithms can be tested under different, controllable conditions. Queries are uniformly sampled from the dataset (sampling query IDs). The candidate document set associated with the query ID is provided by each dataset. Then OLTR algorithms generate a result list of documents to display. Clicks are simulated based on two fixed variables: the click probability and the position bias. 

The click probability is the probability of a user clicking on a document after observing it. This probability is conditioned on the document's relevance label. Following previous OLTR work~\cite{oosterhuis2018differentiable,zhuang2020counterfactual,hofmann2011balancing,hofmann2013reusing,oosterhuis2016probabilistic}, we set two types of click behaviour: \textit{perfect} and \textit{noisy}. The click probability of the \textit{perfect} click behaviour is proportional to the relevance level of the documents, and has $0$ probability for non-relevant documents. This simulates an ideal user that is able to always determine the relevance of a document in the SERP. The \textit{noisy} click behaviour mimics instead a realist behaviour on SERPs by assigning a small click probability to non-relevant documents and a small skip probability to relevant documents. Table~\ref{click_model} provides the click probabilities for the two user models. 

Position bias is modelled by the document observation probabilities; we assume the observation probabilities only depend on the rank position of the document and set these probabilities to:
\begin{equation}
P(o_{k}=1 | k) = \left( \frac{1}{k} \right)^\eta
\end{equation}
where $k$ is the rank position and $\eta$ is a parameter that determines the level of position bias. Following Joachims et al.~\cite{joachims2017unbiased} and Jagerman et al.~\cite{jagerman2019model}, we set $\eta=1$. 
Thus, the probability of a click occurring on a document at rank $k$ in the result list is:
\begin{equation}
P(c_k=1) = P(c_k=1|o_k=1, rel(d_k))P(o_k=1 | k)
\end{equation} 

\newcolumntype{b}{X}
\newcolumntype{s}{>{\hsize=.5\hsize}X}
\newcommand{\tc}[1]{\multicolumn{1}{c}{#1}} 
\setlength{\tabcolsep}{3mm}
\begin{table}
	\centering
	\caption[centre]{Click probabilities for different user behaviours.}\label{click_model}
	\begin{tabularx}{\linewidth}{bsssss}
		\toprule
		&  & \multicolumn{4}{c}{$P(c=1|o=1, rel(d))$} \\ 
		\midrule
		$rel(d)$ & 0& 1 & 2 & 3 & 4\\
		\midrule
		\textit{perfect} & 0.0 & 0.2 & 0.4 & 0.8 & 1.0\\
		\textit{noisy} & 0.4 & 0.6 & 0.7 & 0.8 & 0.9\\
		\bottomrule
	\end{tabularx}
\end{table}

\subsection{Evaluation measures}
We measure effectiveness using standard OLTR evaluation practice, which considers two aspects: \textit{offline} and \textit{online} performance.

\textit{Offline} performance is the final convergence of the learned ranker. We evaluate this using the average nDCG@10 of the ranker over the queries in the held-out test-set across $100,000$ impressions, as all rankers would have reached convergence at this point. 

\textit{Online} performance measures user experience during training. This is quantified by the nDCG@10 obtained by the rank list $L_i$ that the user observes in the training episode $i$, times a discount rate $\tau < 1$: 
\begin{equation}
online\_performance = \sum_{i}\tau^{i}\cdot NDCG(L_i)
\end{equation}
The discount rate assigns less weight to the later impressions to reward OLTR algorithms that learn an effective ranker fast, so to limit the amount of low-quality user experience. As in previous work~\cite{oosterhuis2018differentiable,wang2019variance,zhuang2020counterfactual,hofmann2013reusing}, we set $\tau=0.9995$: this means that impressions that occur after  $10,000$ iterations have less than a $1\%$ impact. We further note that, while this online performance measure aims to quantify the user experience during training, it only does so partially: the relevance of the results in SERP is in fact only one of the many aspects influencing user experience~\cite{al2010review,maxwell2017study}. These other aspects however are not measurable in the context of the typical simulated experiments performed in OLTR research.

\subsection{Experimental runs}
We compare ROLTR to four OLTR baselines. First, the \textit{Dual Bandit Gradient Descent (DBGD)}~\cite{yue2009interactively} method is used as it is one of the standard and most influential algorithms for OLTR. This method uses interleaving for online evaluation, where only one candidate rank is compared to the production ranker at each update step.

The second baseline we consider is the \textit{Probabilistic Multileaving Gradient Descent (PMGD)}~\cite{oosterhuis2016probabilistic}, which has been reported to be the best traditional OLTR method that uses online evaluation with multileaving comparison~\cite{oosterhuis2018differentiable,zhuang2020counterfactual}. For this baseline, we use the same hyper-parameters settings reported in previous work~\cite{zhuang2020counterfactual}, where the number of candidate rankers is $n = 49$, the step size is $\delta =1$  and learning rate is $\alpha = 0.01$.

The third baseline we consider is \textit{Counterfactual Online Learning to Rank (COLTR)}~\cite{zhuang2020counterfactual}, which uses DBGD to update candidate rankers but is combined with counterfactual evaluation. This method has been reported to be more effective than online evaluation based methods in terms of final convergence, but it does present an overall deterioration of user experience during the learning cycle. For this baseline, we use the best hyper-parameters from the original paper~\cite{zhuang2020counterfactual}.

The last baseline we consider is the \textit{Pairwise Differentiable Gradient Descent (PDGD)}~\cite{oosterhuis2018differentiable}, which represents the current state-of-the-art in OLTR. This method does not require online evaluation; instead, it directly optimizes a ranker using gradient descent, and its gradient estimations are unbiased with respect to document pair preferences inferred from user clicks. The comparison between ROLTR and PDGD is interesting because both methods use gradient descent but they differ in the way the unbiased gradient estimation is computed. Following the original PDGD paper~\cite{oosterhuis2018differentiable}, we set its learning rate to $\alpha = 0.1$.

ROLTR has two hyper-parameters: the reward function and the learning rate $\alpha$. We study six reward functions in total: $R_{IPS^+}$, $R_{IPS^-}$, $R_{IPS^+} + R_{IPS^-}$ (i.e. using both clicked and unclicked signal) and their corresponding unshaped naive reward functions. We set the learning rate $\alpha=0.01$ for MSLR-WEB10K and Istella, and $\alpha=0.005$ for Yahoo! (values tuned based on validation set). 

To simulate selection bias, we set $M=10$ to only display 10 documents in the result lists for all experimental runs. For fair comparison, all methods are used to optimize a linear ranker. In order to measure statistically significant differences between methods, all runs are repeated 15 times spread evenly over the available dataset folds with different random seeds. The results are reported and compared using averaged performance and the two-tailed t-test. 

In addition to comparing ROLTR with the other OLTR baselines mentioned above, we also compare our method against the MDPRank trained using full information, i.e., the offline LTR settings with actual relevance labels. This is to be thought of as the skyline for ROLTR as this method is built on MDPRank but training in ROLTR occurs with partial, noisy information (user clicks). For MDPRank, we use the experiment settings from the original paper~\cite{wei2017reinforcement}.

\section{Results}


\subsection{RQ1: Impact of the reward discount factor}
\label{sec_discount}
To answer RQ1, we study the reward discount factor $\gamma$ (set to 1 in previous work~\cite{wei2017reinforcement}). As discussed in Section~\ref{policy_gradient}, we expect that setting $\gamma=0$ will lead to smaller variance in gradient estimation and a consequent speed up of training, without loss in terms of final convergence. 


\begin{figure}
\centering
\begin{subfigure}{.5\textwidth}
	\centering
	\includegraphics[width=1\linewidth]{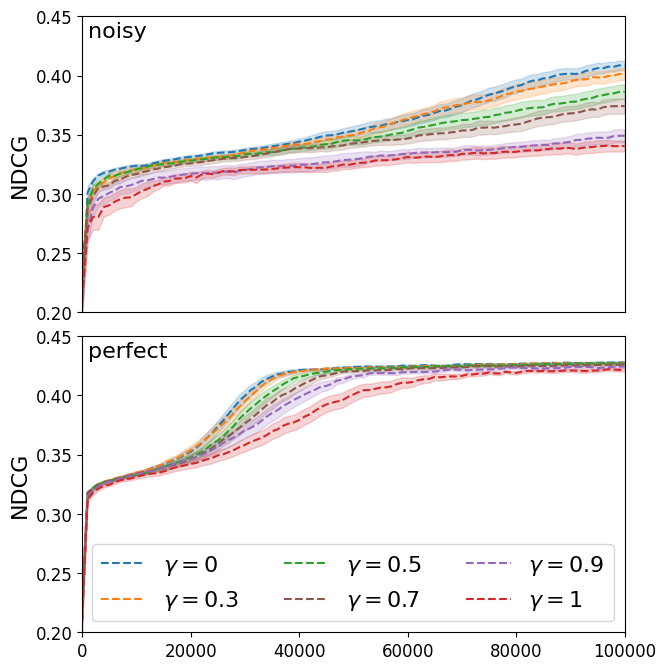}
	\caption{Reward function: $R_{NAIVE^+}$}
	\label{fig:sub1}
\end{subfigure}%
\begin{subfigure}{.5\textwidth}
	\centering
	\includegraphics[width=1\linewidth]{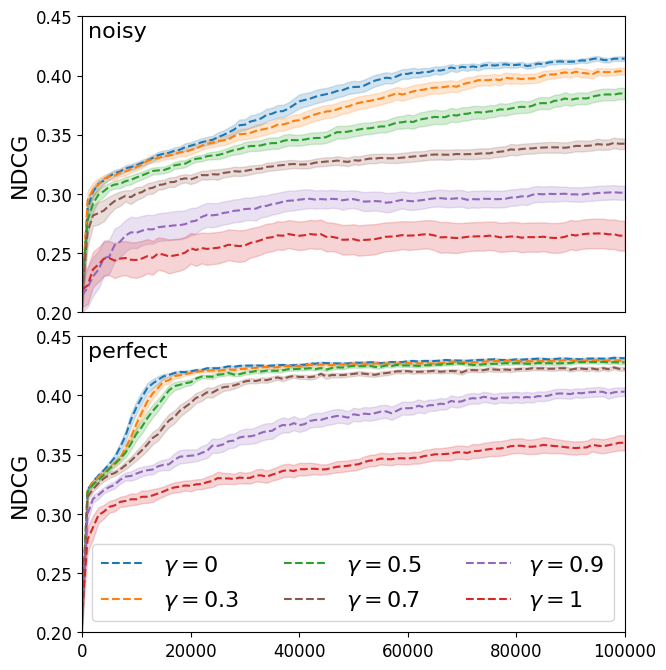}
	\caption{Reward function: $R_{IPS^+}$}
	\label{fig:sub2}
\end{subfigure}
\caption{Offline nDCG@10 score of ROLTR with different reward discount factor $\gamma$. (MSLR10K dataset)}
\label{fig:1}
\end{figure}

\begin{figure}
	\centering
		\includegraphics[width=0.9\linewidth]{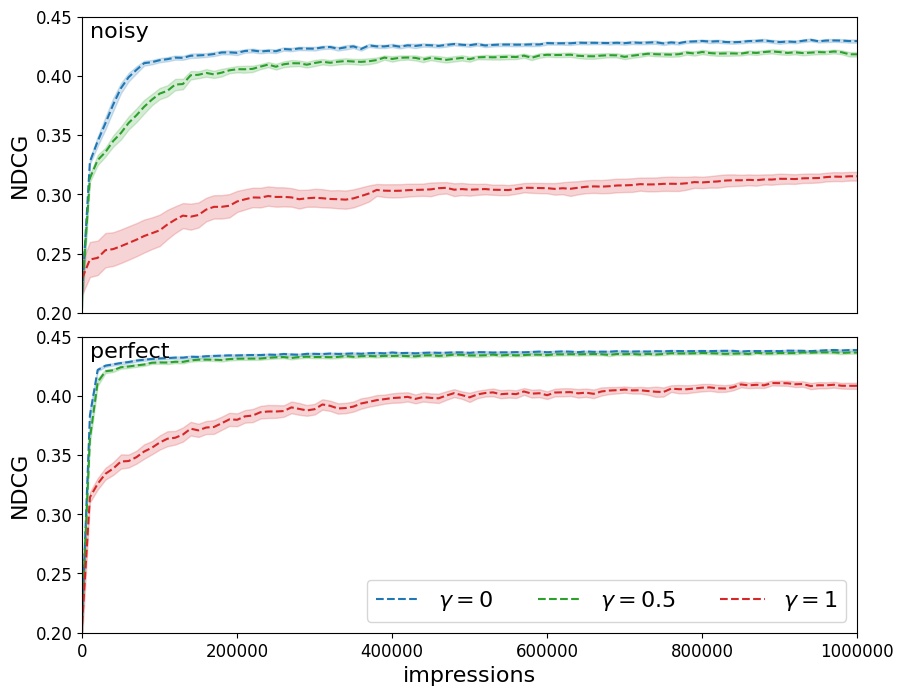}
	\caption{Long-term offline nDCG@10 score of ROLTR with different reward discount factor $\gamma$. (MSLR10K dataset, reward function: $R_{IPS^+}$).}
	\label{fig:7}
\end{figure}

To confirm this assumption, in this section we empirically investigate the convergence of the learned rankers with different values of $\gamma$. In this study, we use the reward function $R_{NAIVE^+}$ and $R_{IPS^+}$, and simulate the user propensity with $\eta=1$. We report results on MSLR10k. Figure~\ref{fig:1} illustrates the offline nDCG@10 learning curves obtained throughout the training process. The plots clearly show that the learning curves obtained when $\gamma=0$ (blue curves) is always above that associated with other $\gamma$ values, for both reward functions and both click settings. For naive reward and noisy click setting, rankers trained with lower value of $\gamma$ achieved higher offline nDCG@10 score at the point of 100,000 impressions. This indicates that ROLTR learns fastest when $\gamma=0$. On the other hand, ROLTR converges slower when $\gamma$ increases, and when $\gamma=1$ (red curves) the convergence speed is the slowest. Interestingly, unbiased reward function $R_{IPS^+}$ seems more sensitive to $\gamma$ than the  naive reward function $R_{NAIVE^+}$. Although $R_{IPS^+}$ has faster learning speed than $R_{NAIVE^+}$ when $\gamma$ is small (e.g., $\gamma = 0, 0.3, 0.5$), the learning speed decreases dramatically when $\gamma$ is large, and $R_{IPS^+}$ even fails to converge at 100,000 impressions under perfect clicks. We also studied the convergences of ROLTR with the $R_{IPS^+}$ reward function and $\gamma = 0, 0.5, 1$ for long-term impressions (1,000,000 impressions). These results are presented in Figure~\ref{fig:7}. Similar to the trends in Figure~\ref{fig:1}, larger $\gamma$ values exhibit slower convergence and fail to converge to the same nDCG value as when $\gamma=0$, even after performing one million impressions. This is due to the IPS significantly enlarging the click signal at lower ranks and this effect being cumulated to the rewards assigned to the clicks at higher ranks with large $\gamma$, thus resulting in considerably larger variance.

\begin{figure}[t]
	\centering
	\includegraphics[width=0.8\textwidth]{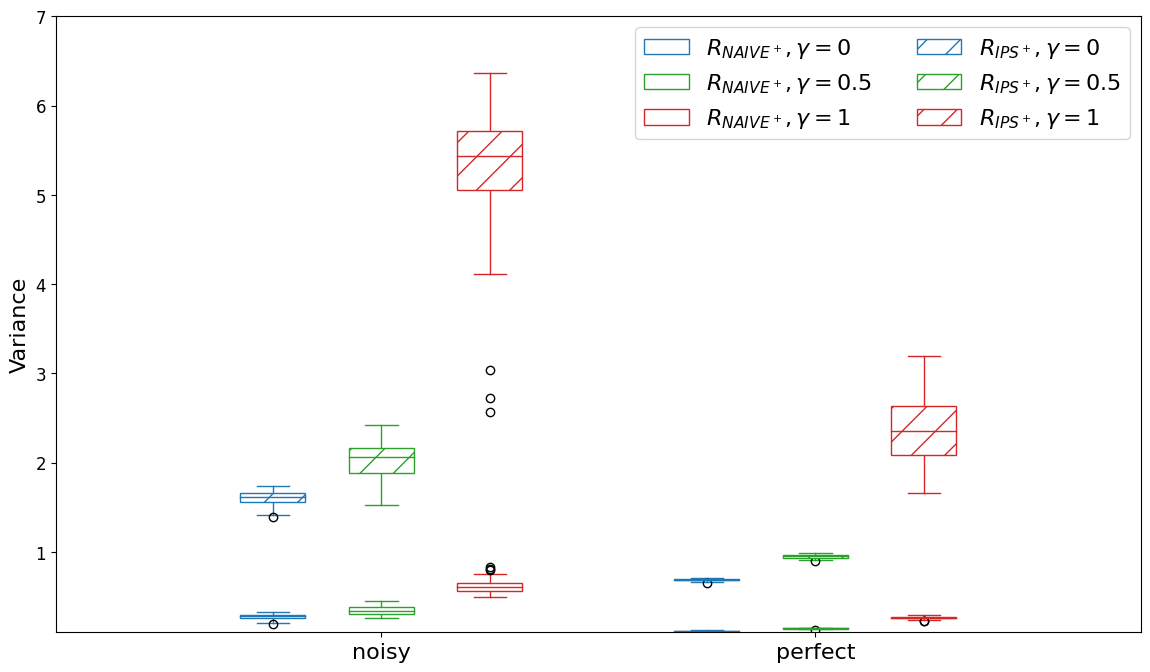}
	\caption{Average variance of the gradient vector for each sampled episode. Error bars correspond to the 95\% confidence intervals. (MSLR10K dataset)}
	\label{fig:2}
	\vspace{-10pt}
\end{figure}

Following Xu et al.~\cite{jun2020ppg}, we also directly compare the variance of the estimated gradients when $\gamma = 0$, $\gamma = 0.5$ and $\gamma = 1$ for both naive and unbiased reward functions. For this, we calculate the variance of the gradient vectors at each training episode using the \textit{trace of the covariance matrix}. We report the results for all runs on the MSLR10K  in Figure~\ref{fig:2} (along to those for $\gamma=0.5$ to show the overall trend for varying values between the two extremes). 
The variance from the noisy click runs is higher than that from the perfect click runs for all values of $\gamma$ and both reward functions. This indicates that strong click noise causes high variance in gradient estimations. Moreover, the variance recorded for $\gamma=0$ is almost half that recorded for $\gamma=1$, for both noisy and perfect clicks and both reward functions. This finding agrees with the reinforcement learning practice: smaller discount factors always result in lower gradient variance. 
When comparing $R_{NAIVE^+}$ and $R_{IPS^+}$, we notice that $R_{IPS^+}$ yields almost 10 times larger variance than that of the same setting for $R_{NAIVE^+}$. This explains why $R_{IPS^+}$ is more sensitive to $\gamma$ and larger $\gamma$ significantly slows down the learning speed, as shown in Figure~\ref{fig:1}. 

In reinforcement learning, it is well known that $\gamma$ controls the trade-off between variance and final convergence. Larger values of $\gamma$ result in higher final convergence at the expense of higher variance (thus slower learning speed), and vice versa: smaller values result in lower variance (faster learning speed) but suboptimal convergence. However, in OLTR, our experiments show that optimizing the expected immediate reward (i.e. $\gamma=0$) will not hurt the final convergence at least after 100,000 impressions: the optimal policy learned with $\gamma=0$ is equal to the optimal policy learned with $\gamma=1$, which is maximizing the DCG score of the ranking. We further note that previous work that formalizing LTR as an MDP  shows high variance in gradient estimation~\cite{wei2017reinforcement} and that recent work has attempted to fix this issue for offline LTR~\cite{jun2020ppg}: setting $\gamma = 0$ can be considered a simpler way to fix this problem. We believe this insight may have positive uptake for offline LTR.

With respect to RQ1, then, setting $\gamma=0$ would not hurt the offline performance (convergence) of the learnt ranker, while delivering lower gradient variance than when $\gamma=1$. Hence, we set $\gamma=0$ in the remaining experiments.

\subsection{RQ2: Impact of reward shaping functions}
\begin{figure}[th]
	\includegraphics[width=\linewidth]{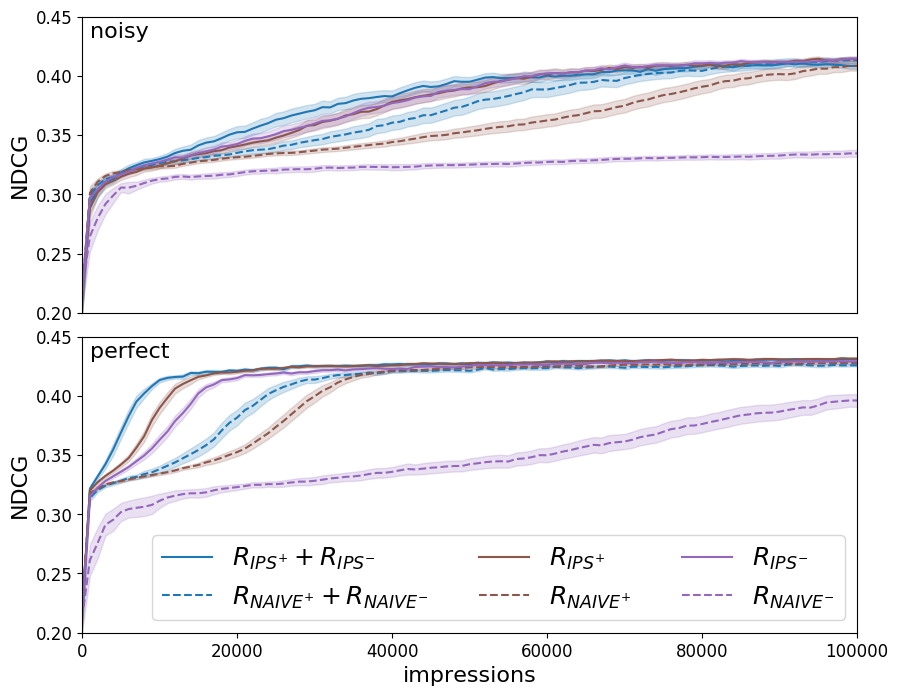}
	\caption{Offline nDCG for ROLTR when different reward shaping functions are used (MSLR10K dataset). }
	\label{fig:3}
	\vspace{-10pt}
\end{figure}

To answer RQ2, we consider the impact of different reward shaping in ROLTR. Figure~\ref{fig:3} reports the \textit{offline} performance of ROLTR with different reward functions over $100,000$ impressions on the MLSR-WEB10K dataset. For both noisy and perfect click settings, all IPS reward functions outperform their naive versions in terms of learning speed. This means that it is important to de-bias the otherwise biased reward shaping function. In addition, we observe that ROLTR with $R_{IPS^{+}} + R_{IPS^{-}}$ and $R_{NAIVE^{+}} + R_{NAIVE^{-}}$   converges faster than when only using positive or negative reward functions. This is under both click settings, and it suggests that leveraging both clicked and unclicked signals is beneficial. This phenomenon is however less obvious with noisy clicks due to noisy clicks decreasing all the rankers' performance, thus resulting in worse and closer nDCG@10 scores. On the other hand, when noisy clicks are considered, $R_{IPS^{+}}$ and $R_{IPS^{-}}$ have a very similar learning curve; while, in the perfect click settings, $R_{IPS^{+}}$ converges much faster than $R_{IPS^{-}}$.
We explain this difference of behaviour between perfect and noisy click settings as follows. In the perfect click setting, $R_{IPS+}$ only uses signal from the clicked documents, which are relevant documents only in this click setting. $R_{IPS-}$ instead also exploits signal from not-clicked documents (it down-weights this signal), which are non-relevant documents only in this click settings -- this however is a worse signal than that from relevant documents only ($R_{IPS+}$). Thus, it is logical to see $R_{IPS+}$ performing better (faster learning curve) than $R_{IPS-}$.
In the noisy click setting, however, the signals exploited by $R_{IPS+}$ and $R_{IPS-}$ are similar: while still $R_{IPS+}$ only uses clicked documents (while $R_{IPS-}$ also uses not-clicked), in the noisy click setting both relevant and non-relevant documents are clicked. This therefore reduces the differences in performance observed between $R_{IPS+}$ and $R_{IPS-}$.

When studying the final convergence at $100,000$ impressions, we do not record significant differences (i.e., $p>0.05$) except for $R_{NAIVE^{-}}$. This exception can be explained as $R_{NAIVE^{-}}$ only assigns biased negative rewards to unclicked documents and as result, in this case ROLTR just aims to avoid selecting documents that give low rewards, but it ignores valuable documents. 

With respect to RQ2, then, we found that using unbiased reward shaping functions to reshape naive rewards make ROLTR converge faster (i.e. it requires less impressions). Specifically, the use of both unbiased positive and negative reward functions make ROLTR converge the fastest. Hence, in the remaining experiments, we set $R_{IPS^{+}} + R_{IPS^{-}}$ as reward function.

The fact that biased rewards do not change the final performance but slow down convergence may seem counter-intuitive at first. 
In fact, in CLTR studies, the biased learning objective results in a local optimal for the final convergence~\cite{joachims2017unbiased}. However, this may not be the case for OLTR. This is because CLTR experiments use a logging ranker to collect clicks and then train a new ranker with the click log: thus the rank of a document is decided by the logging ranker and has been fixed in the log. In OLTR settings like those used here, instead, the deployed ranker is interactively updated after each session: this means every document has a chance to be ranked at top of the displayed rankings hence be observed: these online interventions can eliminate position bias. However, this does not come for free, as it requires a large number of interactions to eventually obtain a good convergence. This is exactly what ROLTR attempts to solve: reduce the number of interactions to reach convergence. ROLTR in fact is the first OLTR algorithm that directly de-biases user position bias, with the results above showing that it can significantly speed up learning (this in turn will translate in better user experience, see Section~\ref{rq4}).

\subsection{RQ3: Final convergence (offline nDCG)}

\begin{figure}[p!]
	\centering
	\includegraphics[width=0.9\columnwidth]{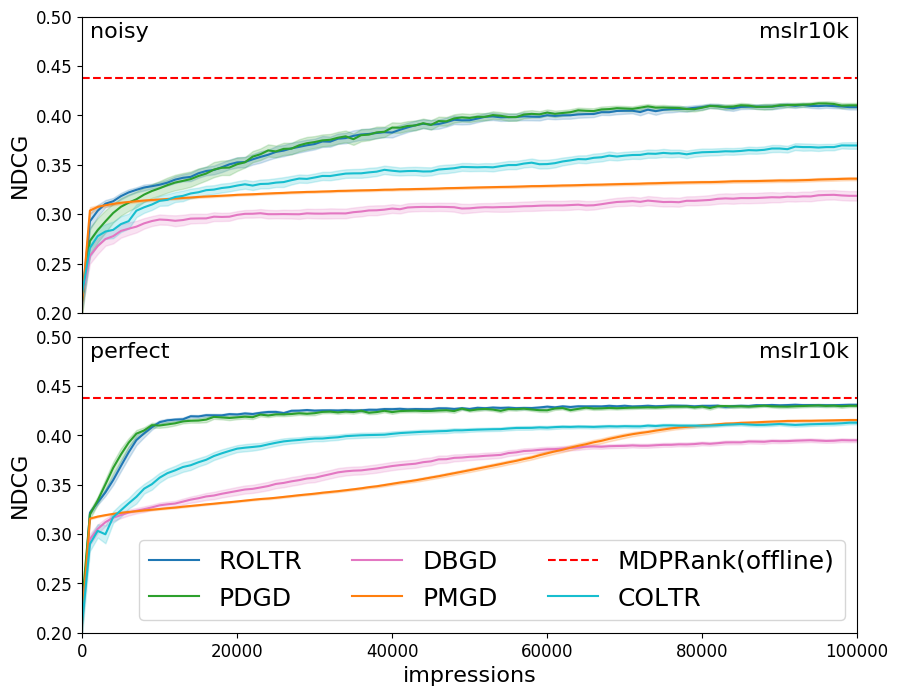}
	\includegraphics[width=0.9\columnwidth]{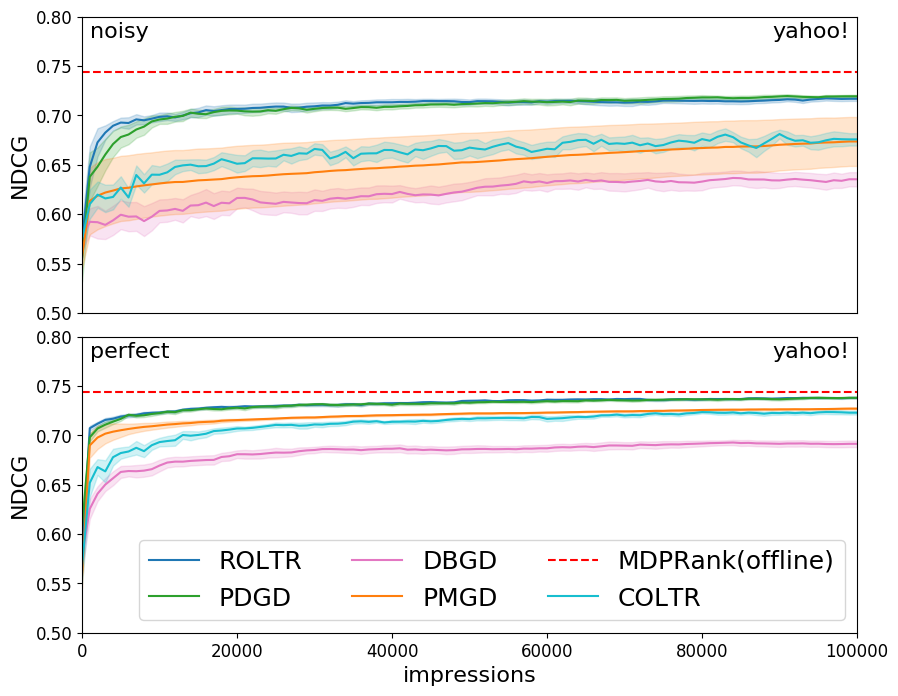}
	\includegraphics[width=0.9\columnwidth]{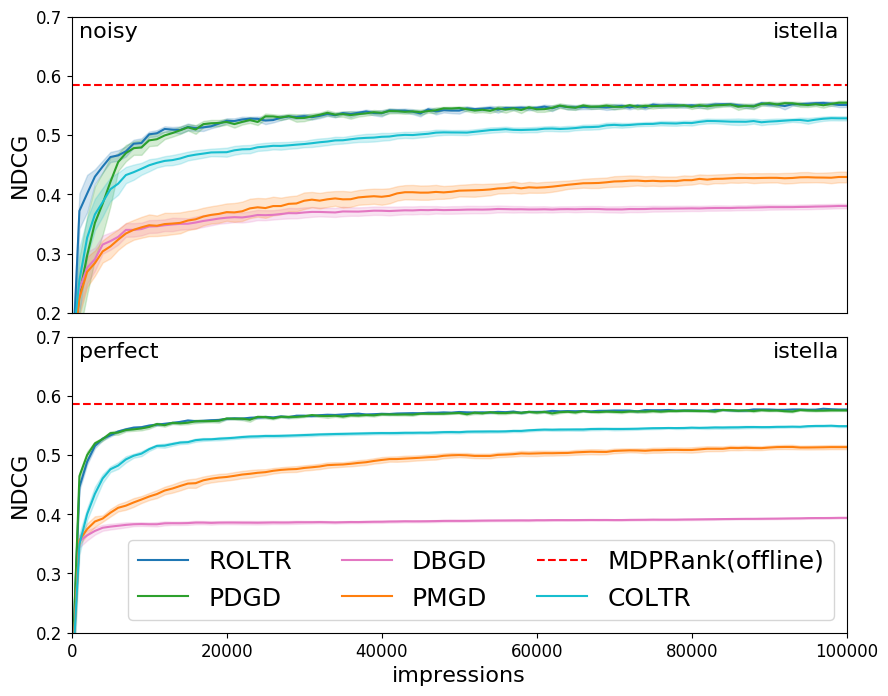}
	\caption{Offline nDCG@10 for OLTR methods and MDPRank under noisy and perfect clicks.}
	\label{fig:4} 
\end{figure}

To answer RQ3, we compare the \textit{offline} performance of the learned rankers over $100,000$ impressions. 
Figure~\ref{fig:4} reports the learning curves for different methods. COLTR is clearly better than PMGD and DBGD in terms of learning speed and final convergence across almost all datasets: this is similar to previous findings~\cite{oosterhuis2016probabilistic,zhuang2020counterfactual}. The gradient descent based methods (ROLTR and PDGD) significantly outperform candidate ranker sampling based methods (DBDG, PMGD and COLTR), rendering DBGD based methods outdated. 

On the other hand, ROLTR and PDGD have similar learning curves across all three datasets. This is especially the case when perfect clicks are used. The final convergence of the two methods across all datasets and click settings is not statistically significantly different ($p>0.05$). However, when noisy clicks are considered, we find that ROLTR has a faster learning speed at the beginning of training. Figure~\ref{fig:5} considers the learning curves for the early impressions (first $10,000$) when noisy clicks are used. This figure clearly shows that the learning curves of ROLTR are almost always above other OLTR baselines at early impressions (except for the first $3,000$ impressions on MSLR10k, for which PMGD is best).

With respect to RQ3, then, we conclude that, when clicks are noisy, the learning speed of ROLTR is faster than PDGD in the early stage of training.

Finally, comparing the final convergence of ROLTR with the effectiveness of the offline MDPRank (trained with relevance labels and thus indicating skyline effectiveness), we can clearly observe that ROLTR's effectiveness is at par to that of MDPRank, when the perfect click setting is used (Figure~\ref{fig:4}). Recall that the perfect click setting still exhibits position bias. This findings empirically demonstrates the unbiasedness of ROLTR. However, when noisy clicks are used for ROLTR, then it's performance is sensibly lower to that of the offline MDPRank, showing how much noise in the click signal hurts performance, compared to perfect relevance labels. 


\newcommand*\rot{\rotatebox{90}}
\newcommand{\B}{\bfseries}
\subsection{RQ4: User experience during training (online nDCG)} \label{rq4}

\begin{figure}[p!]
	\includegraphics[width=\columnwidth]{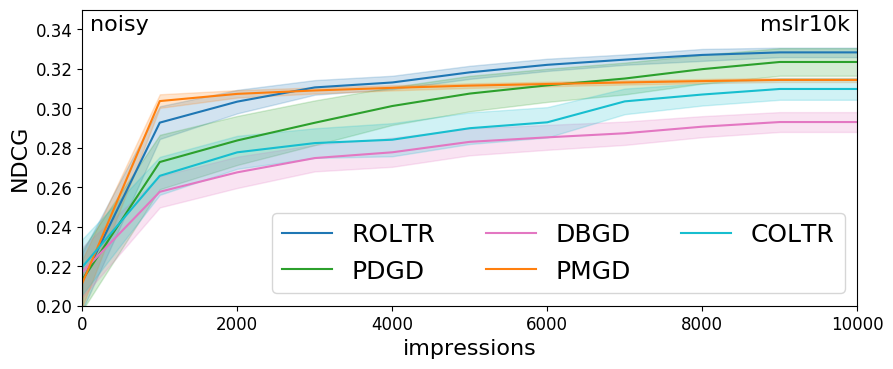}
	\includegraphics[width=\columnwidth]{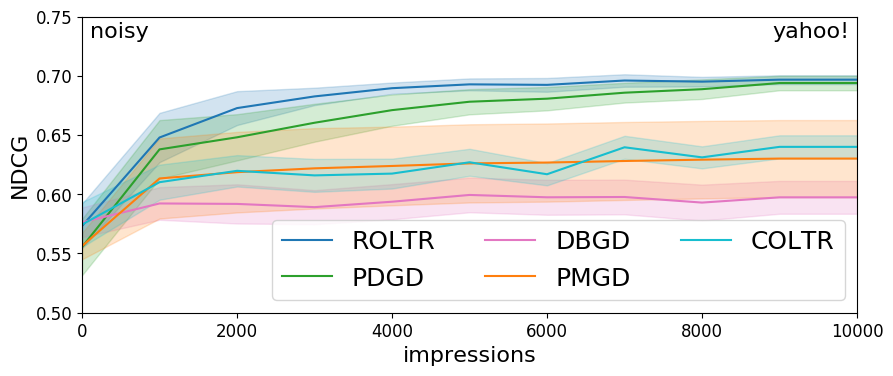}
	\includegraphics[width=\columnwidth]{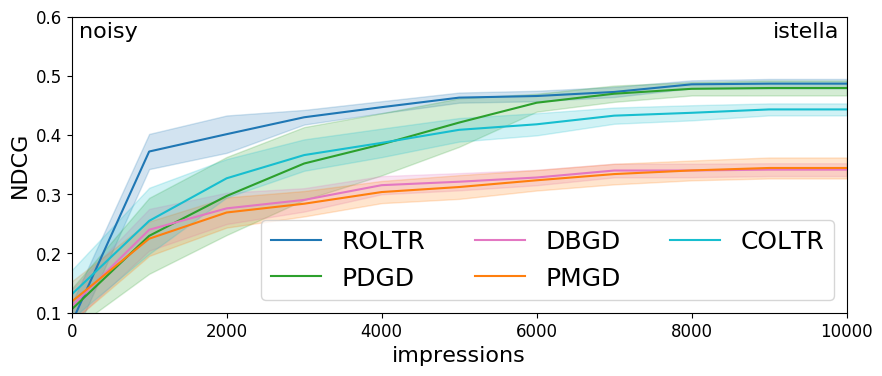}
	\caption{Offline nDCG@10 for the first 10,000 impressions (noisy clicks).}
	\label{fig:5}
\end{figure}

To answer RQ4, we consider the quality of the experience users withstand while rankers are trained (online nDCG). Table~\ref{online} reports the \textit{online} performance of each method. As expected, online NDCG@10 scores obtained when learning from perfect clicks are higher than for noisy clicks, suggesting the latter hurt user experience.
Although COLTR has better offline performance than online evaluation based methods, it does lead to a worse user experience during the learning phase. This indicates that using counterfactual evaluation for candidate ranker comparison requires a lot more exploration. These findings are in agreement with our early findings~\cite{zhuang2020counterfactual}.
We also find that online evaluation based methods are significantly worse than ROLTR and PDGD ($p<0.01$) for both perfect and noisy clicks, agreeing with the offline nDCG learning curves (Figure~\ref{fig:4}). This is because ROLTR and PDGD assemble result lists by sampling documents from a softmax probability distribution, while online evaluation based methods use interleaving or multileaving to create these lists. This suggests that online evaluation based methods perform more exploration, thus hurting user experience more.

When comparing ROLTR and PDGD, we find that ROLTR obtains the best average online performance across all datasets under noisy clicks (agreeing with the learning curves in Figure~\ref{fig:5} for offline nDCG). Even for perfect clicks, for which the offline learning curves of ROLTR and PDGD are similar, ROLTR does statistically significantly outperform PDGD on MSLR10K and Istella ($p<0.05, p<0.01$ respectively). There are no statistically significant differences between ROLTR and PDGD on Yahoo! ($p>0.05$). This may be because the average proportion of irrelevant documents for each query is small (21.92\%), and thus it may be less likely the methods perform a ranking error, making them hard to distinguish. To conclude, with respect to RQ4, we found that ROLTR delivers the best user experience among the investigated methods.

\subsection{RQ5: Sensitivity to propensity mismatch}
\label{sec_propensity_missmatch}

\begin{small}
	\begin{table}[t!]
		\centering
		\caption[centre]{Online nDCG@10. Bold values indicate the highest average performance. Significant gains and losses of ROLTR over PDGD are marked by ${\vartriangle}$ ($p<0.05$) and ${\blacktriangle}$ ($p<0.01$). }\label{online}
		\begin{tabularx}{\linewidth}{@{}p{1.2cm}XXXX@{}}
			\toprule
			Clicks&Algorithm& MSLR10K & Yahoo!& Istella\\
			\midrule
			&DBGD & 519.99& 1165.11&510.34\\
			\multirow{ 2}{*}{\textit{Perfect}}
			&PMGD& 545.22 & 1191.26 &564.58\\
			&COLTR& 448.65 & 1121.92 &348.15\\
			&PDGD & 579.22& \B 1310.99&741.45\\\cline{2-5}
			&ROLTR & \B 587.46 ${\vartriangle}$& 1302.11& \B 808.66${\blacktriangle}$\\
			\midrule
			&DBGD & 477.29&1116.89&408.09\\ 
			\multirow{ 2}{*}{\textit{Noisy}}
			&PMGD  & 535.42  & 1137.39&426.77\\
			&COLTR& 431.27 & 1105.05 &238.30\\
			&PDGD & 516.77& 1227.28&481.85\\\cline{2-5}
			&ROLTR & \B 543.28 ${\blacktriangle}$& \B 1238.10&\B 654.04${\blacktriangle}$\\
			\bottomrule
		\end{tabularx}
	\end{table}
\end{small}

To answer RQ5, we investigate the sensitivity of ROLTR to mismatched user propensity. Our previous experiments, in fact, assumed that the user propensity is known a priori and used the true propensity to obtain the unbiased rewards. However, this assumption does not always hold true in practice, as the user propensity could be overestimated or underestimated for various reasons~\cite{joachims2017unbiased}. 

\begin{figure}
	\includegraphics[width=\linewidth]{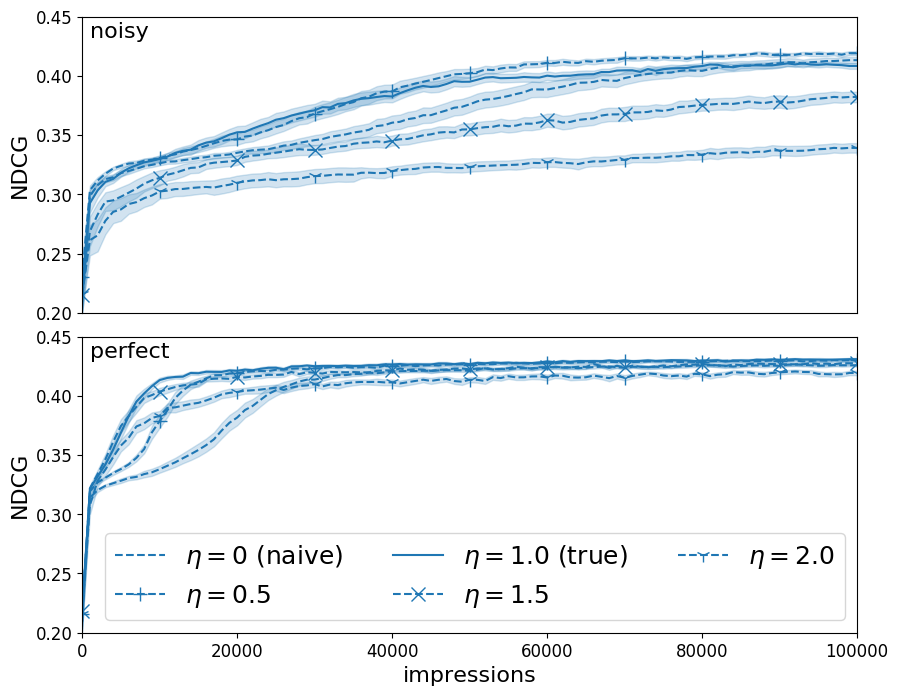}
	\vspace{-6pt}
	\caption{Offline nDCG@10 for ROLTR when different prior propensities are used (MSLR10K dataset).}
	\label{fig:6}
\end{figure}

Figure \ref{fig:6} shows the \textit{offline} nDCG@10 of ROLTR under different propensity values $\eta$ and for different click settings. When $\eta=0$, ROLTR assumes no user position bias and the unbiased reward function $R_{IPS}$ reduces to $R_{NAIVE}$. On the other hand, $\eta=1$ is the same propensity used to simulate clicks, and thus the propensity assumed by the ranker and that seen in the clicks are the same. Note that when simulating the user, we always assume the user has propensity $\eta=1$ (while instead we change ROLTR's propensity value).

From Figure \ref{fig:6} we can observe that, in the noisy click setting, an underestimation of propensity ($\eta<1$) leads to a similar final convergence as that obtained when the true propensity is considered. In fact, when $\eta=0.5$ we record a higher final convergence. However, when the propensity is overestimated ($\eta>1$), both learning speed and final convergence become much worse. A similar behaviour has been found in CLTR and relates to the large variance introduced by the extreme IPS weights~\cite{jagerman2020accelerated}: Because of the overestimated propensities, the IPS weights will be much bigger than those of the true propensity, resulting in a large amount of variance in the gradient estimation, thus hurting the final performance. There are many ways to prevent this from happening, e.g., `propensity clipping' which trades-off bias against variance. 

On the other hand, under perfect clicks, the final convergence of all propensity priors is similar, except for the extremely overestimated propensity ($\eta=2.0$). In fact, in the perfect setting the main difference is with respect to the learning speed: overestimated propensities converge faster than underestimated propensities. 

With respect to RQ5, then, we conclude that, when clicks are perfect, ROLTR is not sensitive to propensity mismatch in terms of final convergence; but when clicks are noisy, only underestimated propensities are not sensitive while overestimated propensities will hurt both the final convergence and learning speed.

\section{Conclusion}
We proposed a novel OLTR algorithm, ROLTR, which formalises OLTR as an MDP ranking problem and uses reinforcement learning with unbiased reward shaping functions to directly optimize an IR metric. In contrast to traditional online evaluation based methods and the current state-of-the-art PDGD method, ROLTR does not fully rely on online interventions to overcome user position bias. Instead, ROLTR directly uses the IPS reward shaping functions to de-bias rewards given by the environment which can be further used to guide gradient estimation. As a result, the gradient calculated by ROLTR is unbiased with respect to position bias. In order to accelerate the learning speed and obtain better user experience, we first simplified the MDP ranking by setting the reward discount factor $\gamma$ to 0, which we empirically confirmed it provides a lower gradient variance without hurting final convergence. Furthermore, to fully leverage user click feedback in each training episode, we introduced a negative IPS reward shaping function for ROLTR which provides additional unbiased reward learning signal from unclicked documents. We have proven that the rewards reshaped by the negative IPS reward shaping function are also unbiased with respect to position bias. Our experimental results show that ROLTR achieves state-of-the-art offline performance requiring less user interactions, which results in considerably better user experience (online performance) over other OLTR methods.

Future work will be directed towards considering estimating user observation propensity during training since our method requires propensity to be known a priori. Furthermore, other biases such as selection bias and other type of user behaviour signals, such as dwell time, mouse move, etc, could be modelled within ROLTR, potentially improving its performance. This could be achieved by simply changing the reward function while maintaining other parts of the method unchanged. This makes ROLTR flexible and with plenty of room for further extension.


%
%

\bibliographystyle{spmpsci}      
\bibliography{ROLTR}


%

\end{document}